\documentclass[sigconf]{acmart}

\usepackage{booktabs} 
\usepackage[ruled]{algorithm2e} 
\usepackage{xspace}
\usepackage{xcolor}
\usepackage{url}
\usepackage{enumitem}
\usepackage{tabularx}
\usepackage{booktabs}
\usepackage{multirow}
\usepackage{ragged2e}
\usepackage{caption}
\usepackage{float}
\usepackage[utf8]{inputenc}
\usepackage{graphicx}
\usepackage{bm}
\usepackage{pifont}
\usepackage{dblfloatfix}

\AtBeginDocument{%
  }

\setcopyright{acmlicensed}
\copyrightyear{2026}
\acmYear{2026}
\setcopyright{cc}
\setcctype{by}
\acmConference[TEI '26]{Twentieth International Conference on Tangible, Embedded, and Embodied Interaction}{March 08--11, 2026}{Chicago, IL, USA}
\acmBooktitle{Twentieth International Conference on Tangible, Embedded, and Embodied Interaction (TEI '26), March 08--11, 2026, Chicago, IL, USA}
\acmDOI{10.1145/3731459.3773333}
\acmISBN{979-8-4007-1868-7/26/03}

\begin{document}

\title{PileUp: A Tufting Approach to Soft, Tactile, and Volumetric E-Textile Interfaces}

\author{Seoyoung Choi}
\affiliation{%
  \institution{Seoul National University}
  \department{Department of Fashion and Textiles}
  \city{Seoul}
  \country{Republic of Korea}
}
\additionalaffiliation{%
  \institution{University of Georgia}
  \department{Department of Textiles, Merchandising and Interiors}
  \city{Athens}
  \state{GA}
  \country{USA}
}
\email{sychoi75@snu.ac.kr}

\author{Rashmi Balegar Mohan}
\affiliation{%
  \institution{University of Georgia}
  \department{Department of Textiles, Merchandising and Interiors}
  \city{Athens}
  \state{GA}
  \country{USA}
}
\email{Rashmi.BalegarMohan@uga.edu}

\author{Heather Jin Hee Kim}
\affiliation{%
  \institution{Cornell University}
  \department{Sibley School of Mechanical and Aerospace Engineering}
  \city{Ithaca}
  \state{NY}
  \country{USA}
}
\email{jk2768@cornell.edu}

\author{Jisoo Ha}
\affiliation{%
  \institution{Seoul National University}
  \department{Department of Fashion and Textiles}
  \city{Seoul}
  \country{Republic of Korea}
}
\email{jisooha@snu.ac.kr}

\author{Jeyeon Jo}
\affiliation{%
  \institution{University of Georgia}
  \department{Department of Textiles, Merchandising and Interiors}
  \city{Athens}
  \state{GA}
  \country{USA}
}
\email{jeyeonjo@uga.edu}

\renewcommand{\shortauthors}{Choi et al.}

\begin{abstract}
  We present PileUp, a tufted pile e-textile sensing approach that offers unique affordances through the tactile expressiveness and richness of its continuous, threaded-volume construction. By integrating conductive yarns in looped or cut pile forms, PileUp transforms soft 3-dimensional textiles into multimodal sensors capable of detecting mechanical deformations such as pressure, bending, and strain, as well as environmental conditions like moisture. We propose a design space that outlines the relationships between texture, form factor, and sensing affordances of tufted textiles. We characterize electrical responses under compression, bending, and strain, reporting sensor behaviors. To demonstrate versatility, we present three application scenarios in which PileUp sensors are seamlessly integrated into soft fabrics: a meditation rug with multi-zone sensing, a fleece sleeve that detects arm motion, and a moisture-sensing wall art. Our results establish tufting as an accessible yet expressive fabrication method for creating integrated sensing textiles, distinguishing our work from traditional flat textile sensors.
\end{abstract}



\begin{teaserfigure}
  \includegraphics[width=\textwidth]{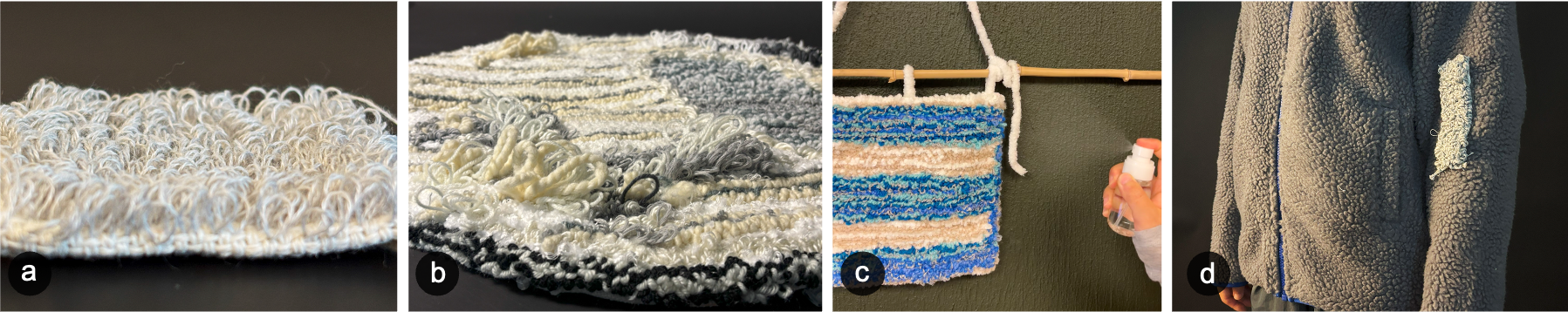}
  \caption{PileUp as tufted e-textile sensors. (a) loop-pile sample made with conductive yarns. (b) meditation rug with multi-zone touch sensing. (c) moisture-sensing wall art. (d) fleece sleeve with bending sensor at the elbow.}
  \label{fig:teaser}
\end{teaserfigure}



\maketitle

\section{Introduction}

Textiles provide a platform for embedding sensing and actuation capabilities into soft, thin, and wearable surfaces. Traditional textile fabrication techniques like weaving, knitting, and embroidery have facilitated the creation of such electronic textiles (e-textiles) with conductive or responsive materials~\cite{guo2023touchandheal, huang2021wovenprobe, luo2021knitui}. These efforts have often prioritized producing forms that are thinner, smoother, and lighter in order to highlight the contrast with conventional rigid electronics, which are characterized by rugged surfaces, lack of conformity, substantial weight, and skin incompatibility~\cite{kim2011epidermalelectronics}. 

While thin and planar forms are emphasized in e-textiles and on-skin applications, the potential for tactile interactions enabled by surfaces with volumes is sometimes overlooked~\cite{ma2022sensequins}. Textiles with volume and rich textures can engage users through enhanced sensory experiences, particularly through touch. Textiles with volume and rich textures can engage users through enhanced sensory experiences, particularly through touch. Volumetric structures enable varied compression and strain responses for richer tactile feedback, support moisture retention for environmental sensing, and create intuitive interaction affordances through their textured surfaces. Tactile communication, as a primary form of human interaction, embodies the transmission of information through haptic feedback, the expression of emotions, and the fostering engagement between individuals and their surroundings~\cite{dellalonga2022affectivetouch, iosifyan2019emotiontextures}. Tactile communication relies on sensory cues, which are derived from texture, temperature, pressure, or vibration~\cite{dahiya2010tactile, hertenstein2009emotiontouch}. Engaging users through carefully designed surface textures is becoming a crucial aspect of fostering user commitment and emotional connection in e-textile development~\cite{delacerda2025tactileemotions, giles2014etextile}.

Pile fabrics, characterized by upright fibers or loops created by tufting, offer three-dimensional features to textile surfaces. Tufted structures are widely used in everyday items such as rugs, terry towels, and velvet for their unique texture and comfort. Prior work has found that tufted textiles encourage users to interact with the surface in specific ways and develop an emotional connection with the product~\cite{pyo2020alltextile}. Despite their potential for unique affordances, they remain largely unexplored in the context of interactive textile research~\cite{goda2024tuftingbags}. These upright fibers in tufted textile create distinctive deformation under compression and shear stress, which could be used to sense touch, stroking, and sliding, when made with conductive fibers. These fibers can also retain moisture, which affects their bulk capacitance, offering another sensing modality. These pile fabrics can also insulate heat for a prolonged period due to their 3D volume, which could be used for therapeutic applications. These unique properties draw distinction from existing flat textile-based interfaces~\cite{slater1977comfort, tanaka2011artificialfurs}.

We present PileUp, a pile fabric structure fabricated with tufted conductive yarns that integrates tactile expressiveness with sensing capabilities (Figure \ref{fig:teaser}). Tufting offers enhanced dimensionality and tactile richness while remaining accessible in terms of tools and materials. We explored how variations in pile structure, pile height, and pile density, as well as yarn properties influence sensor behavior. The sensor’s resistance and capacitance were tested under various conditions, including compression, bending, tensile stretching, and exposure to humidity. Based on these evaluations, we outline a design space for tufted interactive textiles and present three example applications: a moisture-sensing wall art, a fleece jacket, and a meditation rug.

The contributions of this work are as follows:

\begin{itemize}
    \item We introduce PileUp, a novel approach to interactive textiles that leverages tufted pile fabrics with conductive yarns to merge tactile expressiveness and sensing capabilities.
    \item We establish tufting as an accessible fabrication strategy for e-textiles, enabling rich three-dimensional forms and material experiences.
    \item We characterize the performance of  PileUp sensors made with various tufting parameters and yarn characteristics under various deformation modes and exposure to humidity.
    \item We define a design space for tufted interactive textiles and demonstrate its versatility through diverse application scenarios in wearable, domestic, and wellness contexts.
\end{itemize}

\section{Related Work}

\subsection{Tactile Sensing in E-Textile Interfaces}

Tactile sensing in current e-textile interfaces is commonly resistive or capacitive and is largely dominated by planar, sheet-like forms such as woven or knitted textiles with thin and flexible electrodes or laminated layers (KnitUI~\cite{luo2021knitui}, WovenCircuits~\cite{awad2025wovencircuits}, TouchpadAnyWear~\cite{zhao2024touchpadanywear}). These structures offer versatility, light weight, and seamless conformity to a variety of surfaces, in contrast to conventional rigid electronics. Such flat, continuous, and often sealed construction also enhances sensing resolution and reliability by reducing electrical noise from irregular or unpredictable yarn-level contact. However, this design approach often overlooks the rich tactile cues inherent to textile surfaces. Textural interfaces not only create tactile landscapes that invite touch and encourage exploration, but also enable unique sensing behaviors that emerge from yarn interactions. 

E-textiles and on-skin devices with rich textures leverage the textile’s three-dimensional features to enhance user engagement and enable novel sensing behaviors. Embrogami~\cite{jiang2024embrogami} merges embroidery and origami techniques to create foldable, tactilely expressive surfaces that invite manipulation through pressing, folding, and squeezing, implementing fabric-based capacitive sensing. Similarly, spaceR~\cite{aigner2022spacer} extended interaction beyond single touch points, incorporating squeezing, pinching, and whole-hand manipulation to control or navigate content, and relies on a resistive force sensor for input. SenSequins~\cite{ma2022sensequins} introduces playful, touch-responsive sequins that shimmer and react to touch, encouraging exploration through both visual and tactile cues, realizing capacitive touch and on/off contact. Even ultrathin tattoo-like sensors, such as SkinMarks~\cite{weigel2017skinmarks}, can engage microtextures on the skin to anchor touch points and enable intuitive but also engaging input actions via capacitive sensing on conformal skin electronics. Advanced weaving techniques have also demonstrated how layered weaving can deliver complex haptic feedback across larger contact areas, enhancing the sense of material depth~\cite{phillips2005stressrelaxation}.

Pile textiles, characterized by a raised surface of upright loops or cut yarn ends, remain underexplored in e-textile sensing research. The pile fabrics naturally present a soft, three-dimensional surface with distinctive tactile and visual properties that can shape both perception and engagement. Their structure inherently supports intricate yarn-level interactions that could serve as expressive and functional sensing input. ‘Whisker sensors’ with thin, rod-like elements have been explored, but they are designed for robotics and are difficult to integrate into textiles~\cite{yu2024whiskersensor}. Recognizing and leveraging these inherent surface qualities opens opportunities for developing e-textile interfaces that integrate sensing capabilities more deeply with the material’s native structure.  
In this work, we employ pile structures created through tufting, one of the most textural surfaces remaining underexplored in e-textiles, to extend the possibilities for tactile engagement within human–computer interaction (HCI).

\subsection{Structure and Sensing Potential of Pile Textiles}

\begin{figure}
    \centering
    \includegraphics[width=\linewidth]{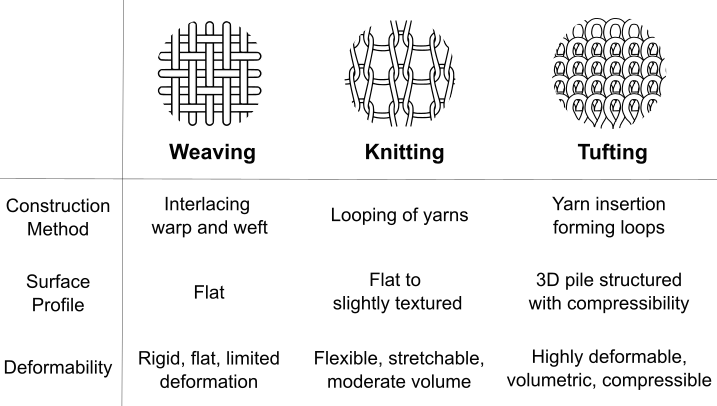}
    \caption{Comparison of common textile fabrication methods, including weaving, knitting, and tufting, showing differences in construction method, surface profile, and deformability.}
    \label{fig:fig2}
\end{figure}

Pile textiles are fabrics distinguished by a raised surface formed from upright loops or cut ends of yarn extending above the woven or knitted base. This pile layer is created during manufacturing by inserting additional yarns into the fabric structure. Originally formed with a continuous yarn, individual piles are laid side by side within a row, and the rows then progress in the direction perpendicular to the pile forming direction. Common examples include carpets, velvet, corduroy, and tufted fabrics. Depending on whether the loops are left intact or cut, the resulting surface exhibits distinct tactile and visual qualities that set pile textiles apart from conventional flat, and planar textiles (Figure \ref{fig:fig2}).

Though other textile fabrication techniques can also achieve surface textures or three-dimensional structures, tufted pile fabrics offer a distinct tactile advantage due to their fully filled, volumetric structure. While knitted or woven structures, as one layer, create a hollow internal form and provide a softer stretch but less substance under pressure. In contrast, tufted piles are composed of densely packed upright yarns that completely fill the fabric volume, resulting in a more cushioned and responsive texture. This dense pile construction provides richer pressure feedback and a sense of depth and softness. Felted structure, on the other hand, can create more discrete three-dimensional structure, but lacks the individuality of fibers or loops. The flat surface of the felted structure provides limited textural variation and responsiveness, while the tufted pile structure stands out for its fullness and multidirectional fiber interaction.

The potential of pile structures in e-textiles has been noted in prior research but remains underexplored in terms of interaction. One noticeable work was a pile-based triboelectric nanogenerator using Ni-coated polyester yarns, demonstrating energy harvesting capability~\cite{pyo2020alltextile}. Studies using the "needle punch" also employed the same pile-forming process as tufting~\cite{jones2021punchsketching}. Jones et al.~\cite{jones2021punchsketching} introduced e-textile crafting using punch needles, yet their focus was on creating soft circuits rather than exploring the sensing capabilities and tactile affordances of pile structures. More recently, tufted structures have been incorporated into design cases for interactive applications. Psarra et al.~\cite{psarra2021sensingtextures} demonstrated tufted pressure and capacitive sensors in tactile textile artworks, while Santos et al.~\cite{santos2025tapestory} created an interactive tapestry using tufted techniques to raise environmental awareness. However, these works focused primarily on application design rather than systematic technical characterization of tufted sensing behaviors. Our work addresses this gap by providing detailed characterization of how tufting parameters influence sensing performance.

Other work has examined the affective and emotional impacts of pile textures in human-robot interaction, such as Puffy~\cite{xing2023puffy} or Touch-and-Heal~\cite{guo2023touchandheal}, though in these cases, the pile textile served primarily as a passive surface covering. PunchPrint~\cite{delvalle2023punchprint}, a 3D-printing method for producing grid-like textile bases, but without extending the approach to interaction using electronics. Early e-textile prior art treated tufted and fuzzy trims like pompoms as capacitive touch electrodes for lighting control~\cite{orth2006touchlight}, providing opportunities for soft tactile interactions, though pile-structure parameters or sensing capabilities beyond capacitive touch were not reported. Across most prior efforts, pile structures have functioned as passive components of an active system rather than being fabricated as the active, interactive medium itself.

The structure of the pile strongly influences texture, durability, and functionality as a sensor. Loop pile, composed of closed, continuous yarn loops, tends to offer greater resilience, elasticity, and wear resistance. Cut pile, in contrast, exposes open yarn ends, producing a softer, plushier surface with increased absorbency. Pile height has a direct impact on mechanical properties such as compressibility and deformation response, which are critical in applications involving tactile sensing or cushioning~\cite{bera2019tuft, phillips2005stressrelaxation}. Longer pile heights increase compressibility but reduce stiffness, while shorter piles provide greater stability and sharper tactile feedback. Tufting is a pile manufacturing technique that allows manipulation of the pile structure. The process works by puncturing the fabric with a hollow or hooked needle, feeding yarn through it, and creating either loop or cut pile structures depending on whether the loops are left intact or severed after formation~\cite{crawshaw1977textile, hartmann1981tuftedcarpeting}.
Our work leverages conductive yarns as both structural and functional elements. The pile itself acts as the sensing medium through inter-fiber contact and structural compression. This approach eliminates the need for additional embedded sensors or layered assemblies, resulting in a soft and volumetric sensing matrix with material and functional integration.

\begin{figure*}[b]
    \centering
    \includegraphics[width=\linewidth]{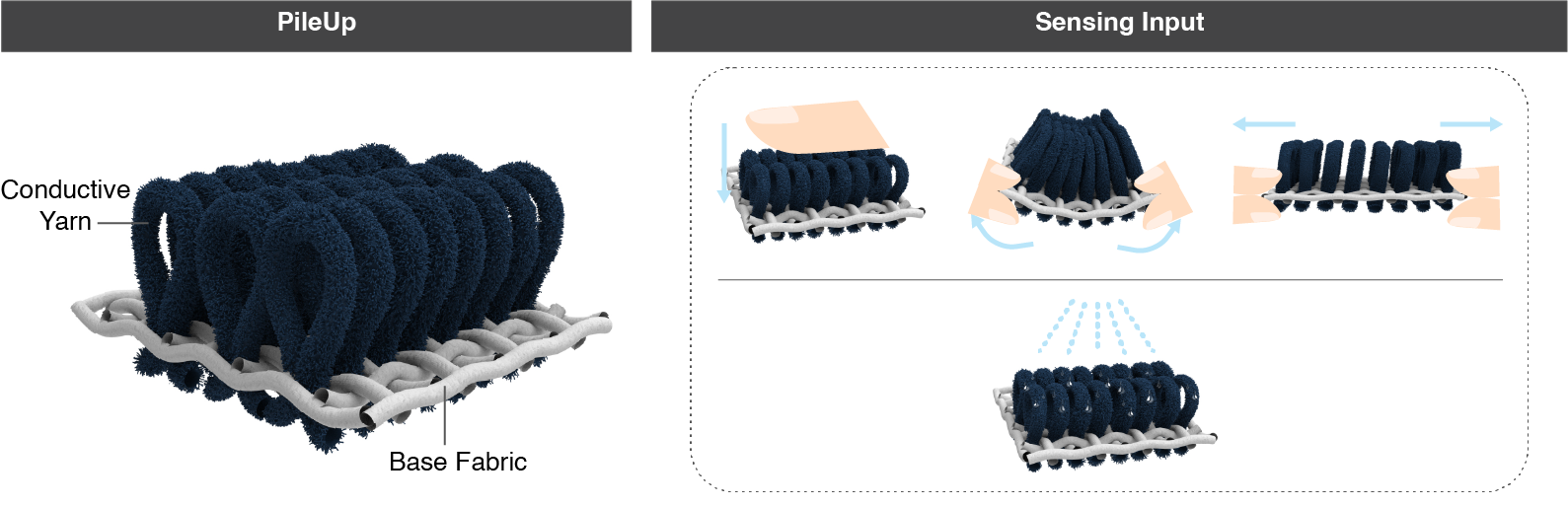}
    \caption{PileUp sensing principle illustrating how tufted conductive yarns detect mechanical deformation (compression, bending, tensile stretch) through resistance change and environmental humidity through capacitance change.}
    \label{fig:fig3}
\end{figure*}

\subsection{Craft-Driven Fabrication as an Interactive Design Practice}

Craft practices have increasingly been recognized in HCI as not just production methods but design approaches that shape how interactive systems are conceived, experienced, and culturally situated~\cite{liu2024hybridcraft}. Textile crafts such as knitting, weaving, embroidery, and felting embody tactile knowledge, material sensitivity, and iterative making that cannot be fully replicated through automated manufacturing. These processes encourage close engagement with the material, supporting designs that can be expressive, context-responsive, and personally relevant~\cite{cheatle2023recollecting, jones2024handspinning}. In the realm of e-textiles, craft-driven fabrication not only influences the physical aesthetics of the interface but also its interactive affordances, as variations in production parameters or material composition can directly alter sensing behavior and tactile experience~\cite{romero2024wovenspeaker}.

Tufting, traditionally viewed as a decorative or craft-based technique, has more recently attracted interest in design research for its potential to support material exploration. It can be understood as a practice that fosters direct engagement with materials and enables creative expression~\cite{lee2023digitaltuftingbee}. Attia and Goda~\cite{goda2024tuftingbags} highlighted the use of tufted surfaces for sustainable design, demonstrating how material reuse and slow processes can foster ecological awareness. Jones et al.~\cite{hertenstein2009emotiontouch} explored the aesthetic and political dimensions of hand-spun yarn, underscoring the tactile and expressive potential of textile crafting as a vehicle for personal and cultural storytelling. This growing body of work positions tufting as not only a method of material construction but also as a socially and culturally embedded design practice. Related craft-based electronic textile literature likewise documents punch-needle techniques in accessibility and domestic making, including needle-punched conductive swatches for touch-based interaction and wall-mounted punch-needle pieces with capacitive sensing~\cite{giles2014etextile, jones2023makingfromhome}.
In this work, we extend the role of craft-driven fabrication by characterizing how tufting parameters influence sensing performance, while allowing the resulting sensors to be integrated seamlessly into textured fabrics. Through characterization, we examined the effects of yarn thickness, pile height, and pile shape (cut versus loop) on sensing performance under different modes of deformation and exposure to humidity. This process transforms tufting from a purely aesthetic or structural choice into a deliberate design space for interaction. By mapping how these material and structural variables shape tactile response and electrical sensitivity, we provide designers with parameters for crafting customized interactive surfaces. This approach empowers practitioners to manipulate tufting parameters intentionally, opening new opportunities for richly expressive, materially grounded interactive textile design.

\begin{figure*}[!b]
    \centering
    \includegraphics[width=\linewidth]{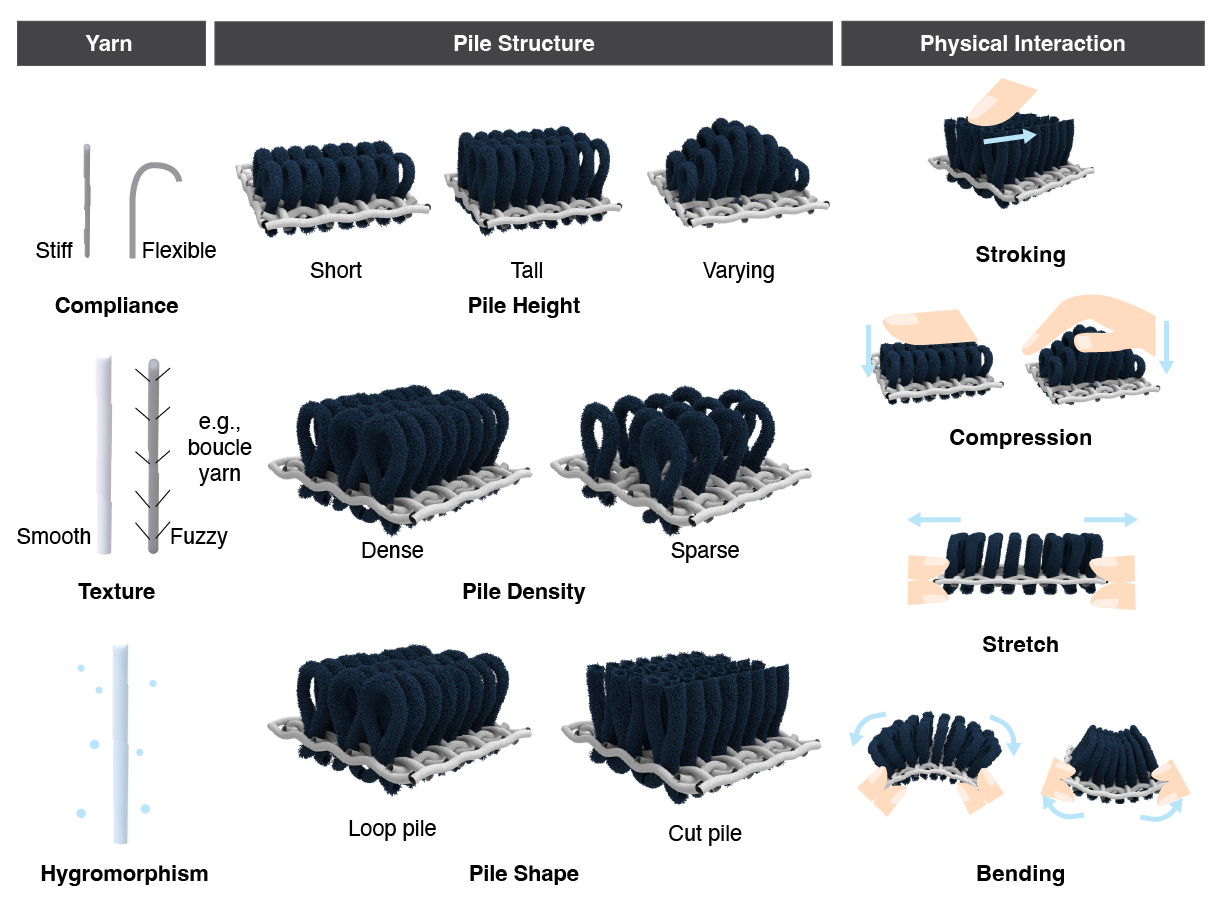}
    \caption{PileUp design space}
    \label{fig:fig4}
\end{figure*}

\section{Design}

\subsection{Sensing Principle}

PileUp leverages dual sensing modalities to detect both mechanical deformation and environmental humidity. It captures mechanical interactions through changes in bulk resistance and senses humidity via capacitance shifts driven by yarn-level dielectric changes. 

PileUp sensor is constructed through the tufting of conductive yarns into a densely packed structure, which is inherently flexible and easily deforms under mechanical stress, such as compression, bending, or tensile stretch (Figure \ref{fig:fig3}). These deformations bring the yarns into close contact, and alter the fabric’s bulk electrical resistance, enabling it to function as a sensor responsive to mechanical inputs. This distinguishes PileUp from other textile-based sensing approaches that use stitched conductive threads, which often avoid contact between threads as they cause short circuits.
At the material level, the use of hygroscopic yarn, such as cotton, introduces an additional layer of sensitivity. These fibers absorb moisture from the environment, leading to changes in the material’s dielectric properties and, consequently, its capacitance ($C = \frac{\varepsilon A}{d}$, where $\varepsilon$ is absolute permittivity, $A$ is the area of a surface, and $d$ is the distance between the surfaces).

\subsubsection{Resistance-Based Sensing of Mechanical Deformation} \hfill

\vspace{.3em}

\noindent \textbf{Compression.} When pressure normal to the PileUp sensor is applied, the pile compresses, increasing the number of contact points between adjacent conductive yarns. This enhanced connectivity creates additional current pathways (i.e., short circuits), leading to a measurable decrease in resistance. The extent of this change depends on parameters such as pile height, contact area, and the mechanical stiffness of the yarns.

\noindent \textbf{Bending.}	Bending introduces asymmetric deformation in the pile structure as a whole. The outer arc (convex side) experiences stretching, while the inner arc (concave side) undergoes compression. In loop pile configurations, bending outward or inward alters the spacing between fibers, modulating electrical conductivity. Depending on the curvature and direction of the arc, resistance can either increase or decrease.

\noindent \textbf{Tensile Stretch.} Stretching the tufted textile increases the distance between loops and elongates the conductive yarns. As the base fabric stretches, the internal geometry of the pile reorganizes, affecting the number, area, and orientation of contact between loops. These microstructural changes influence the continuity of the conductive network, allowing the sensor to respond to directional strain through change in resistance.

\subsubsection{Capacitance-Based Sensing of the Environment.} \hfill

\vspace{.3em}

\noindent \textbf{Humidity.} Unlike resistance, which reflects changes in bulk conductivity, capacitance is determined by the dielectric permittivity ($\varepsilon$) of the material between conductive regions. Water, with a relative permittivity of approximately 78, has a significantly higher dielectric constant than dry textile materials (typically 1.5-2.5). As humidity increases, moisture is absorbed by hygroscopic fibers, resulting in a rise in the local dielectric constant and, consequently, an increase in bulk capacitance. While humidity can also influence resistance through ionic conduction, this effect is typically negligible in comparison to the dominant ohmic conduction through the conductive yarns. Therefore, the primary sensing mechanism under humidity is the change in dielectric permittivity, which drives the increase in capacitance.

\subsection{Design Space}

The PileUp sensor is a multi-modal sensing system that constructs a continuous, three-dimensional volume of yarn (Figure \ref{fig:fig4}). At the yarn level, the physical and mechanical properties of individual yarns, such as permittivity, stiffness, and conductivity, and how different yarns are plied together (i.e., yarn structure) affect not only the sensing performance but also the sensor’s tactile qualities, including texture.

At the structural level, yarns can be tufted in various ways, defined by parameters such as pile height, pile density, pile shape (e.g., loop or cut). Pile height refers to the length of individual piles, while pile density describes the number of piles within a given area. The default pile shape is looped, although these loops can be cut after production to achieve different tactile and visual effects. Loop piles provide continuous electrical pathways, while cut piles rely on fiber-to-fiber contact between densely packed adjacent piles for conductivity. Cut piles offer enhanced softness and sensitivity but with increased signal variance. In a cut pile configuration, when yarns are tufted into the backing at high density, adjacent piles are positioned close enough to physically overlap along their length. This overlap creates multiple contact points between the conductive yarns of neighboring piles, allowing electrical current to flow across the sensor surface despite the loops being cut.

While tufting parameters and yarn properties can influence the bulk electrical resistance of the sensor, their primary impact lies in shaping its affordances. For example, shorter pile heights may invite stroking, whereas taller piles produce a fluffier surface that invites compression. A denser PileUp sensor can lower bulk resistance, thereby improving sensing performance, while also enhancing tactile comfort. Similarly, using yarns with high thermal insulation can produce sensors that feel warmer to the touch.

In the HCI context, these affordances extend beyond functional sensing to encompass tactile and experiential qualities, such as softness, fluffiness, or perceived warmth, that invite particular forms of user engagement. By deliberately tuning these material and structural properties, designers can align the sensor’s physical qualities with specific interaction goals and application scenarios.

\section{Fabrication Process}

\begin{figure}[!b]
    \centering
    \includegraphics[width=\linewidth]{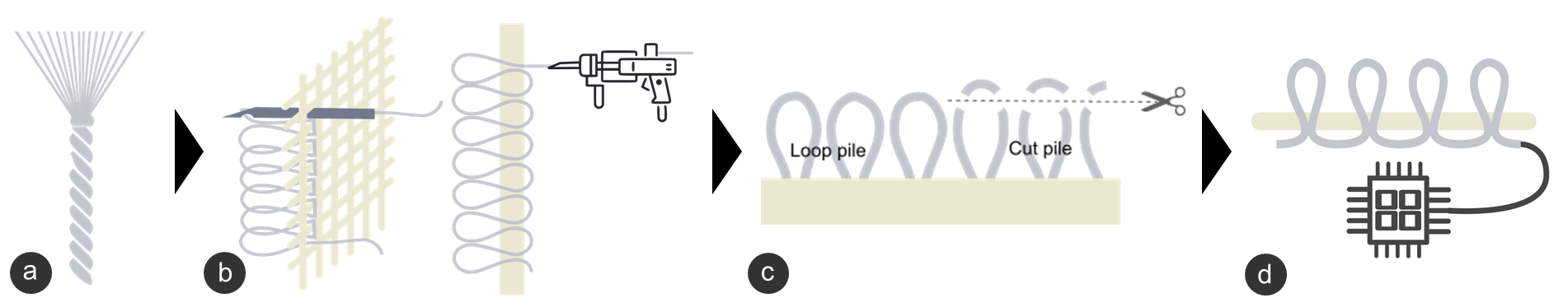}
    \caption{Fabrication process of PileUp sensors. (a) Preparing yarns, (b) tufting onto base fabric using a tufting gun or punch needle (shown vertically to reflect the tufting process, where fabric is mounted in a frame and the tool is inserted perpendicular to the fabric), (c) post-processing to create loop or cut piles, and (d) connecting electronics to the tufted structure.}
    \label{fig:fig5}
\end{figure}

\noindent \textbf{Step 1. Preparing Yarn.} Two approaches are available: use off-the-shelf conductive yarns or customize yarns through plying (Figure \ref{fig:fig5}(a)). Off-the-shelf conductive yarns are usually thin and suitable for manual punch needle tufting. For electric tufting, it is recommended that the commercial yarns be plied. As an empirical guideline, the tufting gun used in our applications (Riiai, China) operated reliably with yarn diameters of approximately $1.5$--$2.0\,\mathrm{mm}$. Therefore, twisting 16 strands of a yarn produced a yarn of about $1.86\,\mathrm{mm}$ diameter with a linear resistance of approximately $30.9\,\Omega/\mathrm{cm}$, which could be fed on its own. An 8-ply configuration measured about $1.1\,\mathrm{mm}$ in diameter and about $14.6\,\Omega/\mathrm{cm}$; it can be co-fed with one strand of non-conductive yarn (e.g., acrylic or wool) to reach a feedable diameter while adding color. Plying can be performed manually or with a basic spinning tool; the target ply count should be adjusted to the specific needle eye and feed characteristics of the tufting gun model, and for the applications, an electric yarn spinner (EEW 6.1, Dreaming Robots, USA) was used. In our work, off-the-shelf yarns were used without modification for creating samples for characterization (Section~\ref{sec:characterization}), whereas customized yarns were prepared by plying multiple strands of a commercial conductive thread to achieve a diameter suitable for an electric tufting gun for our applications (Section~\ref{sec:application}). 

\noindent \textbf{Step 2. Preparing Base Fabric.} Before tufting, the base fabric on which yarn loops are formed is prepared (Figure \ref{fig:fig5}(b)). Monk's cloth is widely used for its durability and structural integrity. However, applications requiring greater stretchability may benefit from alternative base fabrics. The fabric is stretched taut and secured to a wooden frame to ensure consistent tension, an essential condition for precise tufting. Once mounted, the tufting layout is pre-marked directly on the fabric to guide the process and minimize deviations from the intended design.

\noindent \textbf{Step 3. Tufting.} Two primary tufting tools are available for makers: punch needles and electric tufting guns. Punch needles create tufted loops by repeatedly pushing the yarn through fabric with a needle. As the needle moves in and out, it leaves loops of yarn on the other side, building a textured, tufted surface. Punch needles allow for precise control over pile height and stitch placement, making them suitable for detailed work, experimental designs, or delicate adjustments. In contrast, electric tufting guns insert yarns and create loop piles at a high feed rate using motorized needle, which accelerates fabrication for larger uniform regions (Figure \ref{fig:fig5}(b)). However, tufting guns accommodate only sufficiently bulky yarns due to their thicker needle and larger needle eye. They are generally unsuitable for thin yarns, fine details, or adding motifs on already tufted areas, as repeated needle insertions can damage the backing fabric. As an empirical guideline for yarn selection, if the yarn diameter is less than approximately $1.5\,\mathrm{mm}$, a punch needle may be more appropriate, or the yarn can be co-fed with additional strands to reach the required thickness for a tufting gun. The specific diameter threshold depends on the tufting gun model and needle eye dimensions. Depending on application, a tufted sensor with distinct regions can be created with different yarns. Fully conductive zones can be tufted using only conductive yarn, providing areas of high electrical conductivity for sensing. These zones are recommended for areas where reliable sensing is prioritized, such as discrete touch points or regions subject to repeated deformation, where contact among piles needs to be consistently detected. Mixed regions can be created by co-feeding conductive yarn (e.g., 8-ply) with one strand of colored non-conductive yarn. While fully conductive zones offer higher sensing sensitivity, mixed regions serve both functional and aesthetic purposes. In our work, punch needles were used exclusively to control pile height and stitch placement of samples for characterization (Section~\ref{sec:characterization}). For the applications in Section~\ref{sec:application}, both tools were employed depending on the fabrication context: tufting guns were used for larger-scale pieces to accelerate production, while punch needles were reserved for experimental designs (e.g., excessively long piles) that required precise control over unconventional structures.

\noindent \textbf{Step 4. Post Processing.} After tufting, post-processing can be performed to finalize the desired pile shape: loop or cut (Figure \ref{fig:fig5}(c)). This step is critical for tuning the texture, tactile quality, and deformation behavior of the surface. Loop piles, which retain their continuous structure, offer spring-like elasticity, while cut piles provide a softer, plush feel with more flexible fiber ends. Each tufted loop can be manually cut using precision scissors to form uniform cut piles, as demonstrated in S7 in the characterization section (Section~\ref{sec:characterization}). Alternatively, some electric tufting guns are equipped with cutting mechanisms that automatically cut loops during pile formation, which can streamline the process for larger pieces.

\noindent \textbf{Step 5. Connecting PileUp Sensors to Electronics.} After tufting and post-processing, the base fabric is removed from the frame and excess edges are trimmed (Figure \ref{fig:fig5}(d)). Depending on the application, electronic components, including microcontrollers and wiring, can be attached to the back of the textile. Yarns can be connected or soldered to wires to ensure insulation and a secure connection. To ensure a clean finish, edges can be folded and secured. A textile backing can be applied to the tufted side to improve durability and insulate electrical components. This fabrication step can be modified depending on the application and the specific purposes of the sensors.

\section{Characterization}
\label{sec:characterization}

\subsection{Material and Sample Fabrication}
 
\begin{figure}
    \centering
    \includegraphics[width=\linewidth]{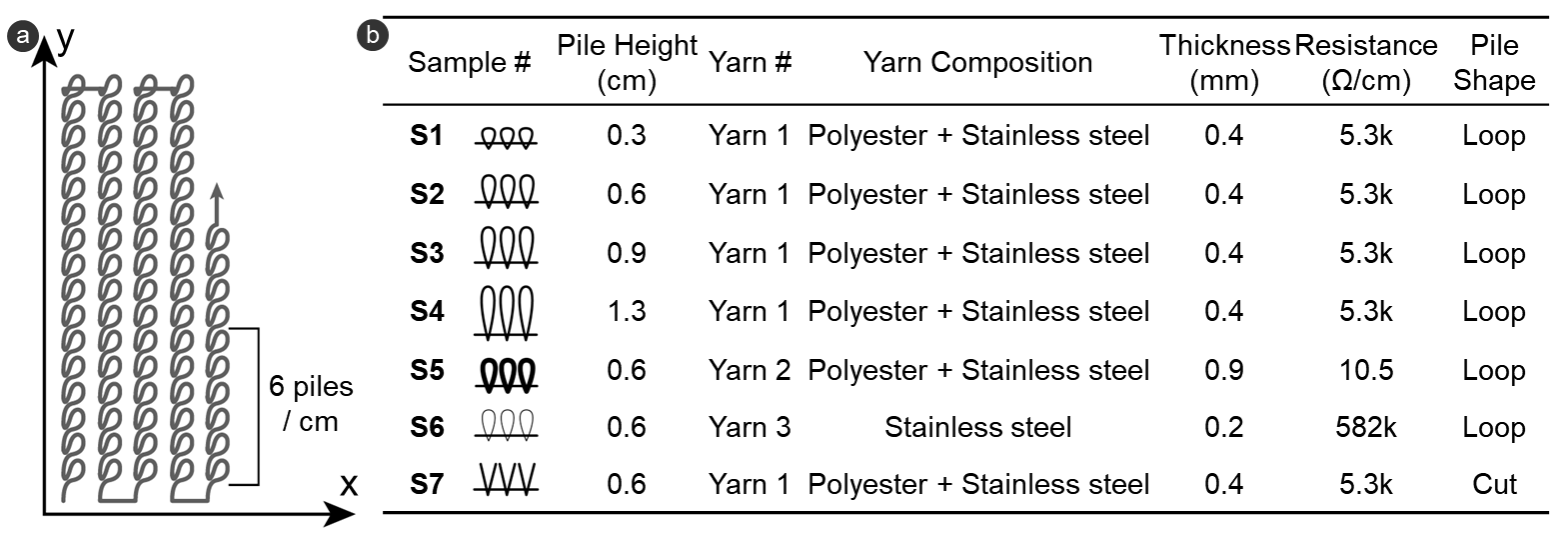}
    \caption{Tufted textile sensor specifications. (a) Pile orientation and stitch density (6 piles/cm). (b) Specifications of the seven sample types (S1–S7) varying in pile height, yarn type, yarn composition, thickness, resistance, and pile shape.}
    \label{fig:fig6}
\end{figure}

To ensure precise control over pile height, density, and yarn thickness, all characterization samples were fabricated using a hand-operated punch needle, using three commercially available conductive yarns (Yarn 1-3) without modification. Punch needles are available in a wide range of sizes that support thin to thick yarns and, although slower, allow accurate control of stitch placement and pile height. Three conductive yarns formed the various tufted pile structure for PileUp (Figure \ref{fig:fig6}(a)): Yarn 1 ($0.4\,\mathrm{mm}$, $5.3\,\mathrm{k}\Omega/\mathrm{cm}$), Yarn 2 ($0.9\,\mathrm{mm}$, $10.5\,\Omega/\mathrm{cm}$), and Yarn 3 ($0.2\,\mathrm{mm}$, $582\,\mathrm{k}\Omega/\mathrm{cm}$). Yarns were tufted onto a $5 \times 5\,\mathrm{cm}^2$ monk’s cloth (i.e., base fabric), which is a cotton woven textile commonly used as a base for tufting, in a single direction, with a consistent stitch density of 6 piles/cm (Figure \ref{fig:fig6}(a)). Pile height was set by the punch needle’s depth setting. Stitch density was controlled by placing one insertion at each intersection of the woven backing’s orthogonal grid, yielding equal spacing along both warp and weft. No trimming or secondary backing was applied after tufting.

As a result, a total of seven sensor types (S1 to S7) were created by varying the pile shape (loop or cut), conductive yarn type (Yarn 1-3), and pile height ($0.3\,\mathrm{cm}$, $0.6\,\mathrm{cm}$, $0.9\,\mathrm{cm}$, and $1.3\,\mathrm{cm}$). Detailed specifications for the sample type are listed in Figure \ref{fig:fig6}(b). Five identical specimens were fabricated for each sensor type for characterization. Sample S2, constructed using Yarn 1 ($0.4\,\mathrm{mm}$ diameter), loop pile, and a pile height of $0.6\,\mathrm{cm}$, served as the standard reference for comparative analysis. Yarn 1, which had a medium diameter among the three options, was chosen as the standard along with a loop pile shape because loop piles reliably maintained electrical connections compared to cut piles, which often cause open circuits. While $0.3\,\mathrm{cm}$ was the minimum reliably achievable pile height, the tufting tool was preset to $0.6\,\mathrm{cm}$ and was therefore used as the standard condition throughout the experiments. Plied yarns that are thick enough for electric tufting guns were not used for characterization because they were too thick for the manual punch needle, which also made it difficult to achieve the same density as with other samples.

\subsection{Evaluation Method}

All tufted textile specimens were conditioned under ambient laboratory conditions prior to testing. In accordance with ASTM D7330-22~\cite{astmD7330} and ASTM D6119-19~\cite{astmD6119}, samples were exposed to $23\,^\circ\mathrm{C}$ and 50\% relative humidity for at least 16 hours without stacking. To reduce surface contamination and minimize pile matting effects, all samples were vacuum-cleaned prior to testing, following the surface reconditioning guidance from ASTM D5252-24~\cite{astmD5252}.

Each experiment was conducted to evaluate how tufted textile sensors respond to three deformations: pressure, bending, and tensile strain, and humidity. For compression, we tested all seven types of samples and compared the results based on pile heights (S1-4 using Yarn 1), yarn thickness (S2, S5, S6 with $0.6\,\mathrm{cm}$ pile height), and pile shapes (S2 loop and S7 cut pile). For bending tests, we focused on curvature radius as the primary variable using a single sample type (S5) that used the thickest yarn among the samples to eliminate confounding factors. For tensile tests, we evaluated directional sensitivity using S2 to focus on the signal changes by strain and stretch directions (x, y, and bias). For humidity tests, we examined the effects of yarn type (Yarn 1-3) and pile shape (loop vs cut) using samples S2, S5, S6, and S7. This approach allowed us to isolate and characterize the effect of each parameter while maintaining experimental efficiency. The specific setup and standards varied depending on the nature of each test, as described below.

\subsubsection{Resistance Change Under Mechanical Deformation}

\begin{figure}
    \centering
    \includegraphics[width=0.6\linewidth]{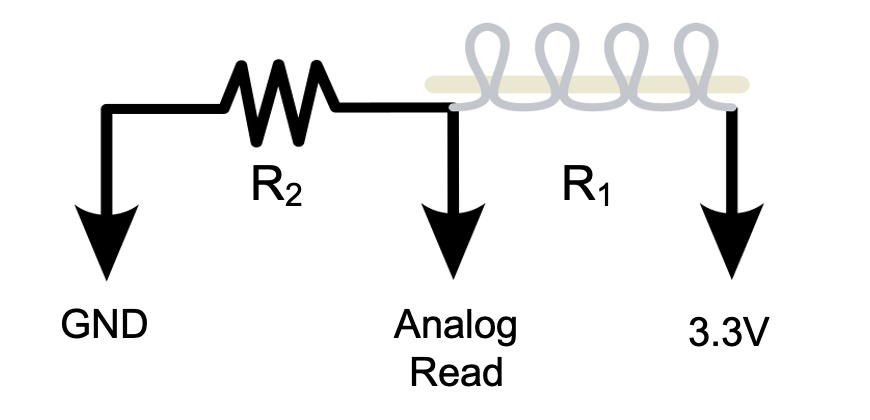}
    \caption{Voltage Divider}
    \label{fig:fig7}
    \vspace{-10pt}
\end{figure}

To examine resistance changes upon mechanical deformation, we connected each sample to a voltage divider circuit (Figure \ref{fig:fig7}) using an analog input pin (A1) of Adafruit Circuit Playground Express (CPX) board ($V_{\mathrm{out}} = \frac{V_{\mathrm{in}} R_2}{R_1 + R_2}$), where $V_{\mathrm{in}}$ is 3.3V from the CPX board, $R_1$ is a PileUp sample as a variable resistor, $R_2$ is a fixed resistor). Normalized voltage change relative to baseline ($\frac{\Delta V}{V_0}$) was used to evaluate the sensor response. Any mechanical deformation that decreases resistance may lead to a higher voltage output and positive normalized voltage changes.

\noindent \textbf{Compression.} To evaluate pressure sensitivity, all sample types (S1–S7) were subjected to a series of controlled compressive loads. Cylindrical metal weights with diameters of 5 cm were incrementally applied to the surface, ranging from $100\,\mathrm{g}$ to $1000\,\mathrm{g}$ in $100\,\mathrm{g}$ intervals. The metal weights were insulated using a dielectric film to prevent contact with the conductive yarns.

\noindent \textbf{Bending.} To assess responsiveness to bending, Sample S5 was wrapped inside/outside of non-conductive cylindrical rods with $1\,\mathrm{cm}$, $3\,\mathrm{cm}$, and $5\,\mathrm{cm}$ diameters. Each bending trial was conducted in both convex (outer) and concave (inner) orientations to determine directional sensitivity.

\noindent \textbf{Tensile Stretch.} The tensile strain response of tufted sensors was evaluated using Sample S2, which was subjected to stretching along three directions: x (perpendicular to the piling direction), y (parallel to the piling direction), and $45^\circ$ bias. For this test, the specimens were fabricated at $75 \times 25\,\mathrm{mm}^2$. Tensile testing was conducted using an Instron universal testing machine, with procedures following ASTM D5034-21~\cite{astmD5034}, which specifies a $75\,\mathrm{mm}$ gauge length, and ASTM D2646-24~\cite{astmD2646}, which sets the test speed at $300\,\mathrm{mm/min} \pm 10\,\mathrm{mm/min}$. Samples were secured in pneumatic grips at both ends and stretched to a maximum displacement of $10\,\mathrm{mm}$ (approximately 13.3\% strain based on the $75\,\mathrm{mm}$ gauge length), with strain calculated automatically by the Instron software as $\varepsilon = \frac{\Delta L}{L_0} \cdot 100\%$. The tensile grip did not touch the conductive yarns.

\subsubsection{Capacitance Change Under Environmental Conditions} \hfill

\vspace{.3em}

\noindent \textbf{Humidity.} Capacitance responsiveness was tested using samples S2, S5, S6, and S7, to determine the effect of the yarn type and pile type. For this test, additional environmental conditioning was applied according to ASTM D1776~\cite{astmD1776}, maintaining $21 \pm 2\,^\circ\mathrm{C}$ and $65 \pm 5\%$ RH. Each sample was sprayed with $5\,\mathrm{mL}$ of water to simulate localized humidity exposure. Rather than using a voltage divider, these tests employed a capacitive sensing electrode of the CPX to detect changes in capacitance of the yarns and surrounding pile as moisture was absorbed. The sensor was connected directly to the capacitive input pin on the CPX board, which allowed for real-time monitoring of capacitance variation without intermediary resistors. Pre-conditioning and test environment were aligned with the same temperature and humidity standards as other tests.

\subsection{Result}

The evaluation results demonstrate that PileUp sensors respond differently depending on yarn properties and pile parameters (e.g., pile height, pile shape).

\subsubsection{Resistance Change Under Mechanical Deformation} \hfill

\vspace{.3em}

\noindent \textbf{Compression.} 
The compression test evaluated voltage changes in response to varying applied weights ($100$--$1000\,\mathrm{g}$, in 100 g increments) using 5 cm-diameter circular weights, while using PileUp 

\begin{figure}[!b]
    \centering
    \includegraphics[width=\linewidth]{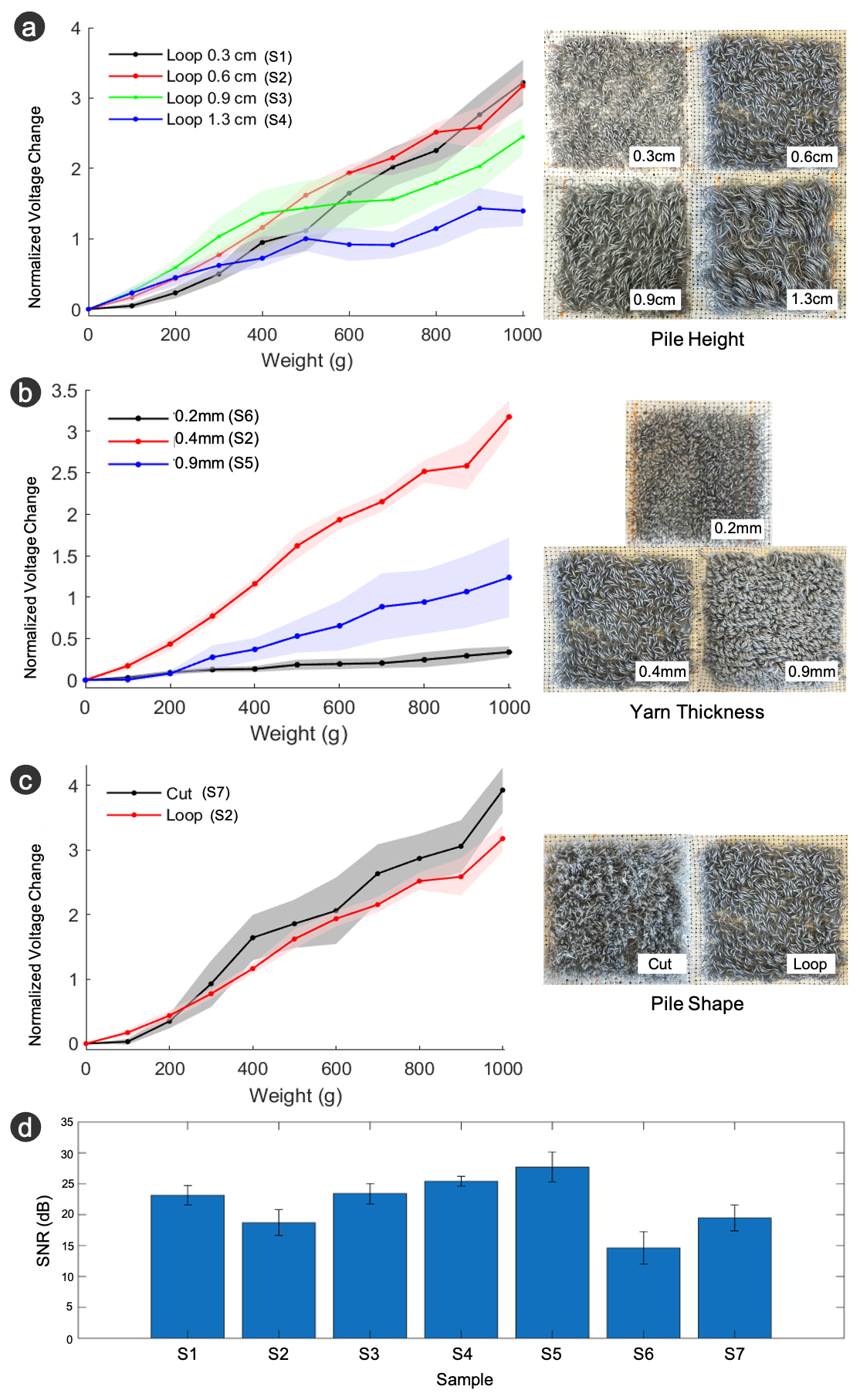}
    \caption{Effect of structural parameters on pressure sensitivity of PileUp sensors. (a) Normalized voltage change for four pile heights (0.3, 0.6, 0.9, 1.3 cm). (b) Normalized voltage change for three yarn thicknesses (0.2, 0.4, 0.9 mm). (c) Normalized voltage change for cut and loop pile shapes. Corresponding sample images are shown on the right. (d) Signal-Noise-Ratio. The error bar represents standard error.}
    \label{fig:fig8}
\end{figure}

samples as variable resistors. Four samples (S1-4) tufted with Yarn 1 were used to compare the effect of the pile heights (Figure \ref{fig:fig8}). Both S1 ($0.3\,\mathrm{cm}$) and S2 ($0.6\,\mathrm{cm}$) responded strongly. As the weight increased, the number of conductive fibers in contact increased, thus decreasing the resistance (i.e., increase in voltage). Long piles did not show many changes, indicating the loops were not that affected by the weight up to $1\,\mathrm{kg}$, because the longer piles formed a stronger bulk stiffness to resist the load. 

When all samples (S2, S5, S6) had identical structure and pile height ($0.6\,\mathrm{cm}$) but varied in yarn type (Yarn 1: $0.4\,\mathrm{mm}$, Yarn 2: $0.9\,\mathrm{mm}$, Yarn 3: $0.2\,\mathrm{mm}$), S2 with Yarn 1 exhibited the highest pressure-sensitivity. The rigid and tightly packed structure of the S5 with Yarn 2 limited deformation, resulting in less change in voltage. Yarn 3 forming S3, being too thin and highly resistive, failed in creating many contacts and generated significantly less voltage changes under pressure.
We compared pile shapes (loop vs cut pile) while holding yarn and the other pile parameters constant. In cut pile configurations, individual fibers are separated from their continuous loop structure, but adjacent piles remain in close proximity for fiber-to-fiber contact. Though the cut pile structure outperformed loops in sensitivity, it sometimes showed unstable responses represented by the larger voltage change variances, as the cut piles behave individually and form new conduction pathways when pressure changes. On the other hand, the piles are all connected in loop structures, so compression increases in normalized voltage change with less variance. All the samples demonstrated a strong signal-noise-ratio (SNR), with a minimum $14.6\,\mathrm{dB}$ of the S6.

\noindent \textbf{Bending.} The bend test examined how curvature affects the resistance of PileUp sensors using S5, made with Yarn 2 at $0.6\,\mathrm{cm}$ pile height and loop structure. The sample was bent along cylindrical surfaces with diameters of $1\,\mathrm{cm}$, $3\,\mathrm{cm}$, and $5\,\mathrm{cm}$, in both concave (inward) and convex (outward) directions.

\begin{figure}
    \centering
    \includegraphics[width=\linewidth]{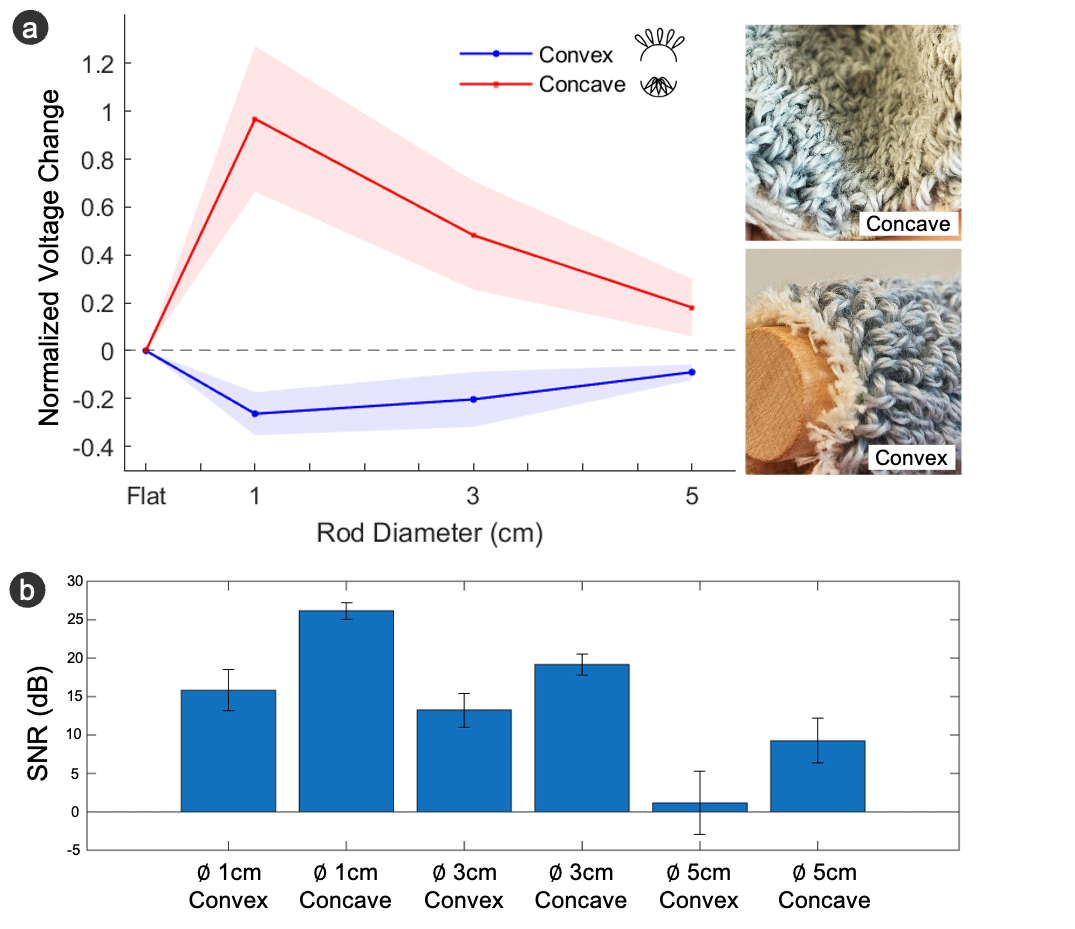}
    \caption{Bending test results for PileUp sensor. (a) Normalized voltage change for convex and concave orientations across rod diameters of 1, 3, and 5 cm. Sample photos on the right illustrate convex and concave bending conditions. (b) Signal-noise-ratio. The error bar represents standard error.}
    \label{fig:fig9}
\end{figure}

When bent inward, the tips of loops were compressed together, increasing inter-fiber contact and resulting in a higher output voltage (Figure \ref{fig:fig9}). When bent outward, the loops were pulled apart, decreasing fiber contact and lowering the signal. Intuitively, the highest curvature of the $1\,\mathrm{cm}$ diameter cylinder produced the largest voltage change in both directions, indicating that tighter curvature caused more significant structural deformation and therefore higher variance in normalized voltage change. As the bending radius increased (diameters $3\,\mathrm{cm}$ and $5\,\mathrm{cm}$), both deformation and voltage change became less pronounced. These results show that loop-structured tufted textiles are responsive to curvature, with output voltage changing depending on bending direction and curvature radius. Yarn flexibility and loop density, influenced here by the thicker Yarn 2 ($0.9\,\mathrm{mm}$), also affect the sensor's performance. As the rod diameter increased, the SNR decreased. The convex bending exhibited a lower SNR compared to the concave bending. For instance, on a 5-cm diameter rod, the SNR was 1.2 dB for convex bending and $9.3\,\mathrm{dB}$ for concave bending.

\begin{figure}
    \centering
    \includegraphics[width=\linewidth]{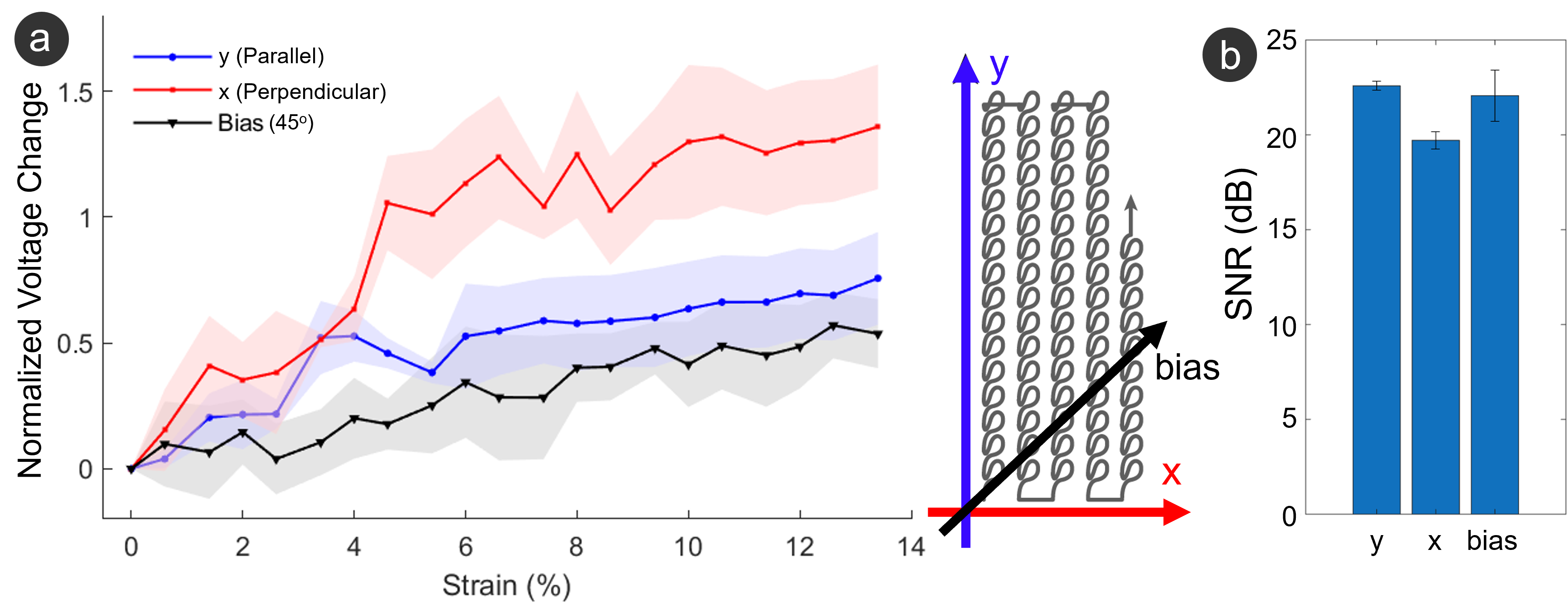}
    \caption{Tensile stretch test results for PileUp sensor. (a) Normalized voltage change under strain applied in three directions: x (perpendicular), y (parallel), and bias (45°) relative to the tufting orientation. Diagram on the right illustrates the three stretch directions. (b) Signal-noise-ratio. The error bar represents standard error.}
    \label{fig:fig10}
\end{figure}

\noindent \textbf{Tensile Stretch.} To evaluate the sensor’s response to mechanical stretching, tensile tests were conducted on S2 (Yarn 1, Loop) in three in-plane directions: x (perpendicular to the tufting direction), y (parallel to the tufting direction), and bias ($45^\circ$) to the tufting orientation. The strain in all directions induced an increase in normalized voltage change (Figure \ref{fig:fig10}), indicating that the base fabric deformed according to Poisson's effect, creating more contacts between the yarns, thus resulting in a short circuit. When strain was applied along the x axis (perpendicular to the tufted row), negative lateral strain caused the yarns in the same row to come closer together and short-circuit. The same occurred when the strain was applied parallel to the tufted rows; however, in this case, the strain along the y axis (parallel to the tufted row) pulled the piles apart in the same row, neutralizing the changes in normalized voltage change. Strain along the bias direction resulted in the least normalized voltage change, implying that the shear deformation caused by bias strain did not allow for larger yarn contacts. The SNR was high for all directions ($> 19.7\,\mathrm{dB}$).

\subsubsection{Capacitance Change Under Environmental Conditions} \hfill

\vspace{.3em}

\noindent \textbf{Humidity.} For the capacitance test, four samples (S2, S5, S6, S7) were evaluated to investigate the effect of yarn structure and pile type on moisture sensitivity. Each sample was sprayed with 
$5\,\mathrm{mL}$ of water, and the change in capacitance was measured relative to its dry baseline. 

\begin{figure}
    \centering
    \includegraphics[width=\linewidth]{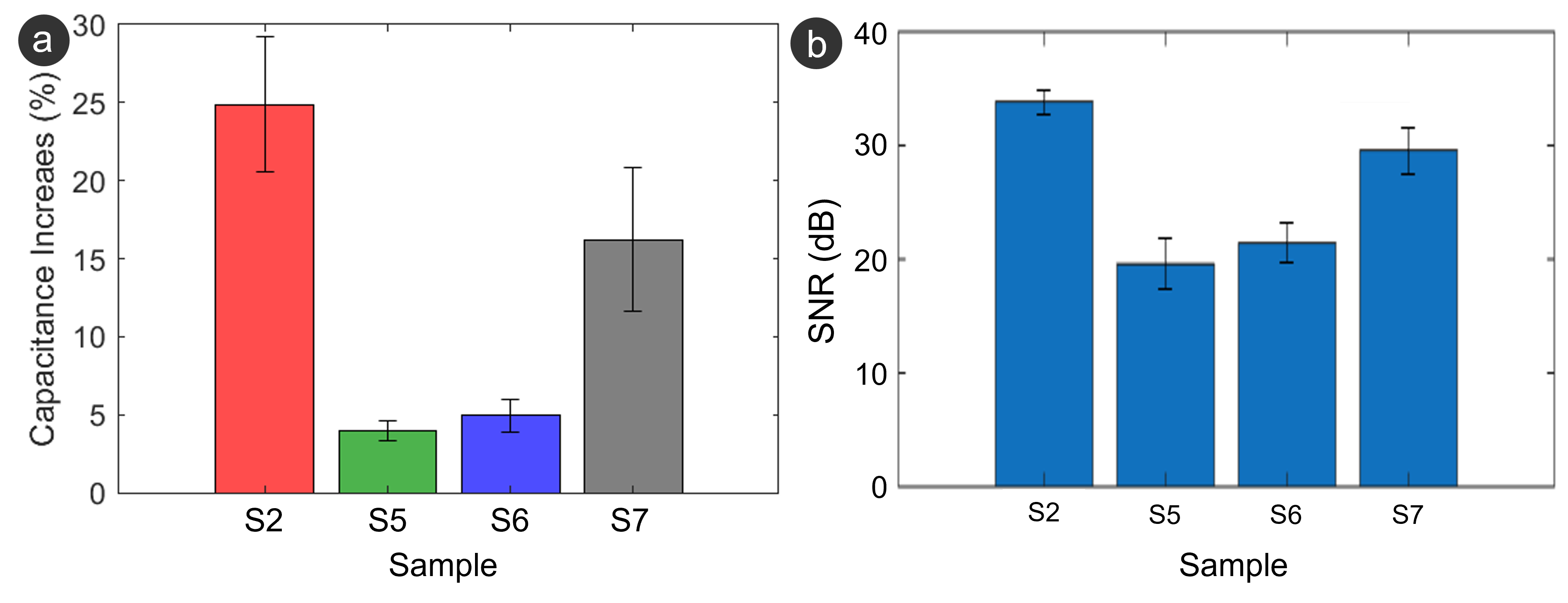}
    \caption{Capacitance test under humidity conditions (a) Change in capacitance for PileUp sensor samples S2, S5, S6, and S7 after spraying 5 ml of water. (b) Signal-noise-ratio. The error bar represents standard error.}
    \label{fig:fig11}
\end{figure}

Across all samples, an increase in capacitance was observed after spraying, suggesting that water infiltration altered the permittivity of the tufted structures (Figure \ref{fig:fig11}). The S2 made of Yarn 1 in loop piles exhibited the largest increase in capacitance. With the density of piles consistent, the thickness differences among Yarn 1 ($0.4\,\mathrm{mm}$), Yarn 2 ($0.9\,\mathrm{mm}$) and Yarn 3 ($0.2\,\mathrm{mm}$) resulted in differences in bulk density. This structure allowed the added moisture to be part of the circuit quickly, thereby increasing capacitance. The S7 made of Yarn 1 with cut piles showed a higher increase in capacitance compared to the other yarn samples, yet a lower increase compared to the loop pile, which indicates that some of the cut piles are disconnected, unlike the continuously connected loop piles. The S5 made of thick Yarn 2 ($0.9\,\mathrm{mm}$) recorded the weakest capacitance change. The thick, densely packed fibers in this sample likely prevented water from deeply penetrating the pile structure, thus the impact of the added moisture was limited. Lastly, the S6 made of thin Yarn 3 also exhibited a slight increase in capacitance. Though its sparse structure allowed moisture to pass through, the thin loop piles had a limited ability to retain water and produce additional capacity to hold the electrical charge. SNR was strong enough in all samples ($> 19.6\,\mathrm{dB}$).

\section{Application}
\label{sec:application}

\subsection{Meditation Rug: Multimodal Interaction for Sensory Mindfulness}

We developed an interactive meditation rug that supports mindful practice through touch sensing and heat feedback (Figure \ref{fig:fig12}(a)). The rug was tufted using Yarn 1 (from Section~\ref{sec:characterization}) with the electric tufting gun, different zones offering specific interaction types. In tufting the rug, we sectioned it into two types of regions: sensing zones and a hybrid zone. The sensing zones, which include the center compression sensor, heater trigger pad at the front right, and three light buttons at the rear right, were tufted exclusively with Yarn 1 in 16-ply to ensure reliable inter-pile contact under repeated use. Any non-conductive colored piles were tufted in separate passes from the conductive ones, ensuring that conductive pathways in the sensing zones rely solely on 16-ply Yarn 1. The hybrid zone was the capacitive touch interface within the front-left haptic zone, where we mixed 8-ply Yarn 1 with a single strand of colored, non-conductive yarn for clearer affordances and richer texture.

The central zone was tufted with 16-ply yarn to serve as a compression sensor that activates the entire system when a user sits on it. Underneath the central zone's piles, we attached a soft heating mat made of carbon fibers enclosed by nonwoven layers. This heater can be activated by a capacitive touch sensor located on the front right side of the rug. The middle picture in Figure \ref{fig:fig12}(a) shows an infrared camera image of the increased ambient temperature around the heating module. On the rear right side, three capacitive buttons turn on the LED lights embedded on the underside of the rug. These buttons not only turn on the lights, but also adjust the brightness to three different levels. Lastly, we embedded a haptic zone on the front left side with long loop piles made of chunky cotton/polyester yarns and Yarn 1 for capacitive touch sensing. These extended piles were created using a punch needle, as the tufting gun could not accommodate such pile heights. Users can find comfort by touching, stroking, or rubbing this zone during meditation, which activates a small vibration motor underneath the piles, enhancing the tactile sensation. The remaining areas of the rug were tufted with non-conductive yarns for decorative purposes, creating visual patterns and borders. For the haptic zone specifically, we used a mixed configuration (8-ply conductive yarn fed together with colored yarn) to provide clear affordances and richer tactile qualities while maintaining sensing functionality. 

This rug demonstrates how textile-based sensors can be integrated into multisensory products that support well-being and embodied interaction. We focused on making each interaction feel smooth, intuitive, and calming.

\subsection{Fleece Jacket Sleeve: Embedding Bending Sensors for Elbow Joints}

\begin{figure*}
    \centering
    \includegraphics[width=\textwidth]{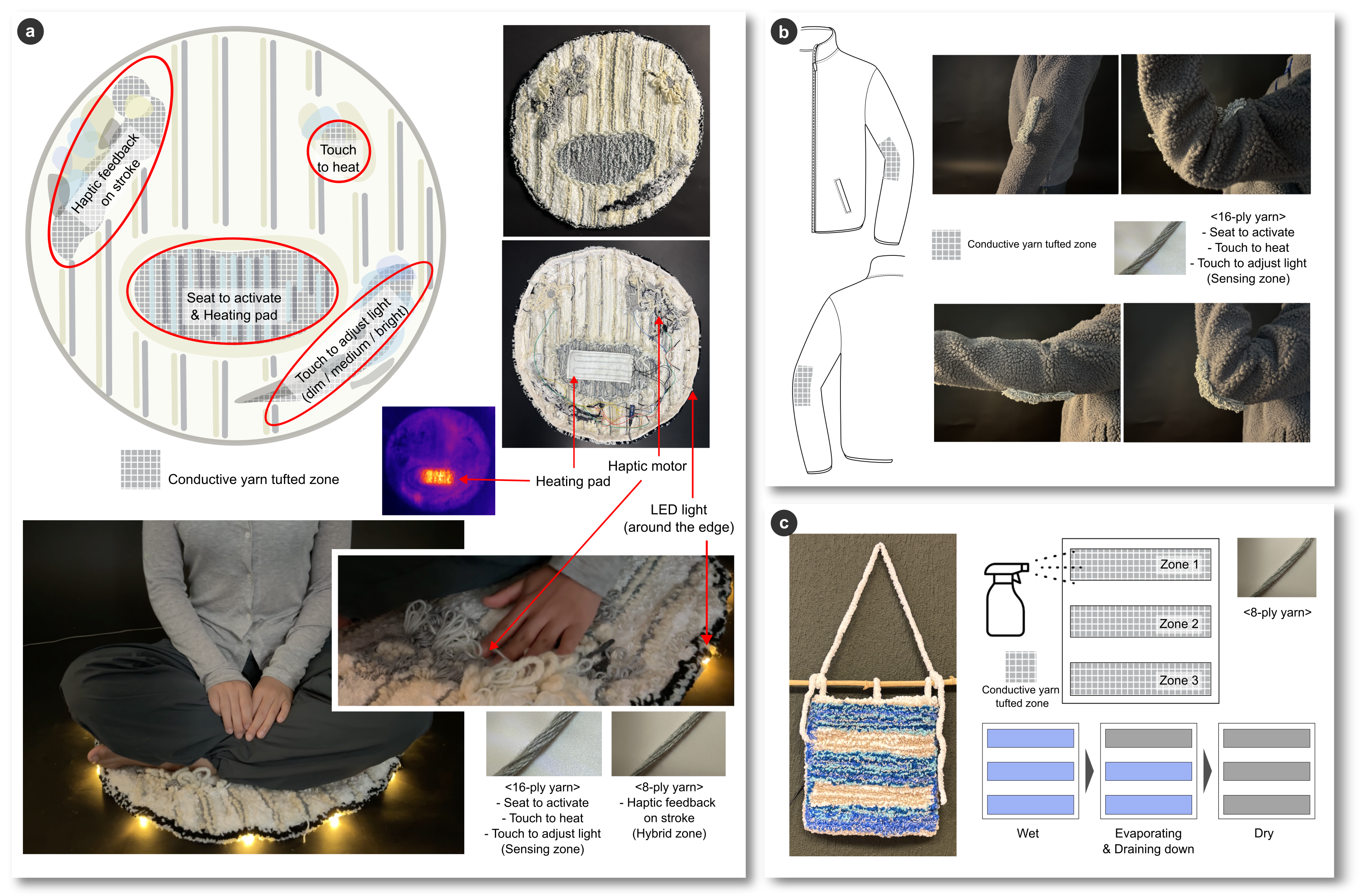}
    \caption{Application examples of PileUp sensors. (a) Meditation rug with multi-zone interaction for pressure, touch-to-heat, and haptic feedback. (b) Fleece jacket sleeve with tufted bending sensor for detecting elbow motion. (c) Moisture-sensing wall art with three sensing zones for visualizing water distribution and evaporation over time.}
    \label{fig:fig12}
\end{figure*}

We applied bending sensors to the elbow areas of a fleece jacket to explore wearable sensing (Figure \ref{fig:fig12}(b)). We tufted loop-structured 16-ply Yarn 1 into the inner and outer elbow sections using the electric tufting gun. The fleece piles on the target area of the sleeve were completely trimmed, leaving only the base fabric. This allows the PileUp piece to be attached using an adhesive. As the arm bends or straightens, the tufted zones experience shape changes that lead to measurable variations in resistance. 
This design shows how tufted textiles can be used as embedded sensors that detect body movement. Extending this application, one could apply PileUp sensors to track physical activities, posture monitoring, and interactive garment interfaces. PileUp sensors blend in naturally with the fleece material, thanks to its textures. 

\subsection{Moisture-Sensing Wall Art: Visualizing Absorption through Conductive Tufting}

In this application, we leveraged the hygroscopic properties of yarns to create a subtle, time-based interaction using capacitance change (Figure \ref{fig:fig12}(c)). The prototype consists of a wall-hanging art type textile sensor composed of three humidity-sensing sections, tufted by feeding 8-ply spun Yarn 1 together with colored yarn through the electric tufting gun.  Upon dampening and suspending the wall art, moisture gradually migrated downward due to gravity. This caused the upper section to dry first, while the lower sections remained wet. The lower sections exhibited increased capacitance, showing the color changes on a connected display. As the wall art continued to dry, the middle section eventually dried out, resulting in decreased capacitance. This process enabled a visual encoding of drying progression over time.

\section{Discussion}

\subsection{Challenges} \hfill

\noindent \textbf{Sensor Performance Variability.} The behavior of individual piles presents a challenge, as they produce noise. Each pile strand moves independently, so repeated or random mechanical inputs rarely deform the surface in the same way. Cut piles make this unpredictable aspect more noticeable, increasing signal variation during characterization tests. While cut piles maintain electrical connectivity through fiber-to-fiber contact between adjacent piles, this configuration does not guarantee consistent conductivity in the way that continuous loop piles do. Individual fibers can separate during interaction, particularly with thinner yarns or lower pile densities, leading to intermittent connectivity. Additionally, the impact of yarn thickness based on plies or the mixed use of conductive and non-conductive yarns on sensor performance have not been characterized. Future research should investigate design parameters such as tuft density, yarn stiffness, and backing materials to improve pile shape retention and enhance sensing consistency.

\noindent \textbf{Real World Applicability.} Our characterization was conducted under controlled laboratory conditions with standardized preconditioning and testing protocols. This approach was necessary to establish baseline sensing behaviors and to enable systematic comparisons across tufting parameters. However, these conditions do not capture the variability of real-world use, where piles may be imperfectly aligned, environmental conditions fluctuate, and user interactions vary in speed, force, and angle. Our application prototypes (the meditation rug, fleece sleeve, and moisture-sensing wall art) demonstrated functional sensing in practical, non-laboratory settings during informal trials. Nevertheless, we did not conduct a systematic evaluation under uncontrolled conditions. The controlled characterization establishes how design parameters map to sensing performance, but the extent to which these findings translate to noisier real-world scenarios remains to be quantified. Although the volumetric structure of tufted piles suggests inherent robustness by relying on aggregate contacts among densely packed fibers rather than precise positioning of individual yarns, this claim requires further validation. Future work will include controlled studies that introduce realistic noise factors such as varied initial pile states, environmental fluctuations, and unconstrained user interactions, as well as long-term reliability testing in deployed applications.

\subsection{Opportunities} \hfill

\noindent \textbf{Embracing Variability and Enhancing Accuracy through Design Aid Tools.} While tufting offers a valuable method for creating craft-based and expressive e-textile fabrication, the manual nature of the process may inevitably introduce performance variability. Future work could deploy design aid tools such as those demonstrated in digital knitting~\cite{albaugh2019digital} to reduce inconsistency while still preserving the expressive and improvisational qualities of hand-made fabrication. This will expand our work to computational craft literature in HCI, in which similar tools have been explored to balance creative freedom and fabrication precision~\cite{buechley2012crafting, devendorf2016postanthropocentric}. We envision future design aid tools could enable hybrid workflows in tufting where the designer visualizes expected sensor performance, pile density, or conductive path connectivity before fabrication, while still leaving space for situated adjustments during making, aligning with HCI discussions around material agency, and human-machine collaboration.

\noindent \textbf{Achieving High Resolution Sensors.} 
 In the current work, relatively large sensing points were fabricated due to limitations in available tools and materials. However, sensing resolution could be substantially improved with industrial machinery specialized for pile textile manufacturing. Such equipment could produce fine pile textiles such as velvet, corduroy, or faux fur that seamlessly incorporate sensing capabilities while preserving their soft and inviting tactile qualities. This would open possibilities for high-resolution touch interfaces, gesture-based controls, interactive furniture, responsive garments, and immersive haptic environments. Integrating sensing in this way could transform everyday textile products into intelligent, responsive surfaces suitable for smart homes, wearable technology, automotive interiors, healthcare monitoring, and adaptive environmental control systems.
 
 \noindent \textbf{Potential for Integrated Actuation System.}  While this study emphasized conductive yarns for sensing, the tufted pile structure also presents an ideal platform for integrating fiber-based actuators or responsive materials. Shape memory alloys or polymers, optical fibers, and heating elements can be embedded into the pile to enable active functions such as shape morphing, localized heating, light emission, or haptic feedback. The vertical orientation and flexibility of pile fibers allow these actuators to be integrated discreetly, maintaining softness, comfort, and aesthetic appeal. This capability points toward multifunctional pile fabrics that can both sense and respond to environmental or user inputs, enabling applications in adaptive fashion, reconfigurable upholstery, interactive art, therapeutic textiles, and soft robotics.

 \noindent \textbf{Tufting as an Expressive Form of E-Textile Sensing.} Unlike common e-textile fabrication techniques such as weaving or knitting, which offer limited spontaneity during the making process, tufting allows the maker to deliberately modify parameters in real time through direct bodily engagement between the material and the tool. The maker maintains continuous physical contact with the textile substrate, receiving immediate tactile and visual feedback that enables exploratory adjustments to pile height, density, or yarn configuration as the textile takes shape. While this work primarily focuses on PileUp’s functionality as a novel subset of e-textile sensors, future work could shift the emphasis toward the maker’s perspective to explore the expressive potential of the tufting process. A design workshop with e-textile practitioners could further this exploration by comparing fabrication methods and fostering new discourse around material expression in e-textile making.

\section{Conclusion}

In this paper, we introduce PileUp, a multi-modal fabric sensor that leverages the tactile expressiveness of tufted textiles. Tufting, an accessible and structurally versatile fabrication method, enables the construction of sensors capable of detecting compression, bending, tensile strain, and humidity, all while maintaining a rich, texturized form. We explore how variations in tufting parameters, such as pile height, type, and shape, as well as yarn characteristics, influence sensor performance. We present both a fabrication process and a design space to support further exploration and adaptation of PileUp. Notably, the three-dimensional volumetric texture of PileUp invites behavioral affordances, offering a departure from conventional fabric-based sensors, where texture is often treated as noise and minimized. By actively utilizing the expressive qualities of tufted loops, we demonstrated three application scenarios in which PileUp sensors are seamlessly integrated into fabrics and garments. Each application detects tactile input, body motion, or humidity, and engages users through outputs such as warmth or color – interactions made possible by PileUp’s unique, fabric-like textured form.

\bibliographystyle{ACM-Reference-Format}
\bibliography{bibliography}

\end{document}